\documentstyle[epsfig,here,12pt]{article}
\begin{document}

\newcommand {\ber} {\begin{eqnarray*}}
\newcommand {\eer} {\end{eqnarray*}}
\newcommand {\bea} {\begin{eqnarray}}
\newcommand {\eea} {\end{eqnarray}}
\newcommand {\beq} {\begin{equation}}
\newcommand {\eeq} {\end{equation}}
\newcommand {\state} [1] {\mid \! \! {#1} \rangleg}
\newcommand {\sym} {$SY\! M_2\ $}
\newcommand {\eqref} [1] {(\ref {#1})}
\newcommand{\preprint}[1]{\begin{table}[t] 
           \begin{flushright}               
           \begin{large}{#1}\end{large}     
           \end{flushright}                 
           \end{table}}                     
\newcommand{\drawsquare}[2]{\hbox{%
\rule{#2pt}{#1pt}\hskip-#2pt
\rule{#1pt}{#2pt}\hskip-#1pt
\rule[#1pt]{#1pt}{#2pt}}\rule[#1pt]{#2pt}{#2pt}\hskip-#2pt
\rule{#2pt}{#1pt}}

\newcommand{\Yfund}{\raisebox{-.5pt}{\drawsquare{6.5}{0.4}}}
\def\Acknowledgements{\bigskip  \bigskip {\begin{center} \begin{large}
             \bf ACKNOWLEDGMENTS \end{large}\end{center}}}

\newcommand{\half} {{1\over {\sqrt2}}}
\newcommand{\dx} {\partial _1}

\def\Dslash{\not{\hbox{\kern-4pt $D$}}}
\def\cmp#1{{\it Comm. Math. Phys.} {\bf #1}}
\def\cqg#1{{\it Class. Quantum Grav.} {\bf #1}}
\def\pl#1{{\it Phys. Lett.} {\bf #1B}}
\def\prl#1{{\it Phys. Rev. Lett.} {\bf #1}}
\def\prd#1{{\it Phys. Rev.} {\bf D#1}}
\def\prr#1{{\it Phys. Rev.} {\bf #1}}
\def\np#1{{\it Nucl. Phys.} {\bf B#1}}
\def\ncim#1{{\it Nuovo Cimento} {\bf #1}}
\def\lnc#1{{\it Lett. Nuovo Cim.} {\bf #1}}
\def\jmath#1{{\it J. Math. Phys.} {\bf #1}}
\def\mpl#1{{\it Mod. Phys. Lett.}{\bf A#1}}
\def\jmp#1{{\it J. Mod. Phys.}{\bf A#1}}
\def\aop#1{{\it Ann. Phys.} {\bf #1}}
\def\mycomm#1{\hfill\break{\tt #1}\hfill\break}

\begin{titlepage}
\titlepage
\rightline{TAUP-2480-98}
\rightline{\today}
\vskip 1cm
\centerline{{\Large \bf Comments on (Non-)Chiral Gauge Theories}}
\centerline{{\Large \bf and Type IIB Branes}}
\vskip 1cm

\centerline{Adi Armoni\footnote{armoni@post.tau.ac.il} and Andreas 
Brandhuber\footnote{andreasb@post.tau.ac.il}}

\vskip 1cm
\begin{center}
\em School of Physics and Astronomy
\\Beverly and Raymond Sackler Faculty of Exact Sciences
\\Tel Aviv University, Ramat Aviv, 69978, Israel
\end{center}
\vskip 1cm
\begin{abstract}
We use type IIB brane configurations which were recently suggested by
Hanany and Zaffaroni to study four dimensional 
N=1 supersymmetric gauge theories. We calculate the one loop beta
function and realize Seiberg's duality using a particular
configuration. We also comment on the anomaly cancelation 
condition in the case of chiral theories and the beta function in the 
case of chiral and SO/Sp theories. 
\end{abstract}
\end{titlepage}
\newpage

\section{Introduction}

In the last two years we have learned that branes are a useful tool
in the study of supersymmetric gauge theories \cite{HW}(for a recent
review and a complete list of references see \cite{GK}).
Branes give a nice and effective pictorial meaning to symmetries and
parameters of the low energy field theory.

Recently, there has been progress in the realization of chiral 
theories using branes \cite{LLL,BHK,EGKT,HZ} of type II string 
theory. In particular,  
Hanany and Zaffaroni \cite{HZ} constructed various classical N=1 
chiral four dimensional field theories, using brane configurations of
type IIB string theory, in the limit of zero string coupling.

In this note we use the suggested construction in the presence
of non-zero string coupling. In this case the fivebranes will bend,
with asymptotic bending which is dictated by supersymmetry \cite{AH,AHK}.
The bending of a (p,q) fivebrane which is stretched in a line in the
(x,y) plane is 
\beq
 \Delta x : \Delta y = p + q\tau,
\eeq
where $\tau={i\over g_s} + {\chi \over 2\pi}$ and $g_s,\chi$ are the
string coupling and type IIB axion. Moreover, it was explained in
ref. \cite{GG} that the slope of each brane should remain constant
along the line even if the brane crosses other fivebranes. This constraint
leads to anomaly cancelation of local symmetries in the field theory. 

The paper is organized as follows: In section 2 we use the bending of
the branes to calculate the beta function of N=1 supersymmetric QCD. 
In section 3 we suggest a brane configuration and brane moves which
reproduce Seiberg's duality. 
In section 4 we introduce a configuration for chiral gauge theories
which is determined by anomaly cancelation in the field theory and
bending constraints similar to those in \cite{GG}.
We also give an approximate calculation of the beta function in these
cases. We comment on the construction of brane configurations with 
symplectic and orthogonal groups in section 5.

Some of the issues that we discuss were previously addressed in the 
framework of type IIA string theory and M-theory
\cite{EGK,EGKRS,HOO,Witten1,BIKSY}. The issue of finite gauge
theories, using type IIB branes was discussed recently in \cite{HSU}.  

\section{N=1 SQCD setup and the beta function}

The brane configuration that we use is similar to the one that was
introduced in ref.\cite{HZ}. The configuration (Fig. 1) consists of two
horizontal lines which represent two NS
branes with world volume 012367  (which we will denote by NS') 
and two vertical lines which represent NS
branes with world volume 012345 (which we will denote by NS).
In addition, there are D5 brane
with world volume 012346 which are attached to the NS and NS' branes. 
This set of branes leads to N=1 SQCD in four dimensions with
generically chiral matter. 
The numbers in the boxes of Fig.1
correspond to the number of D5 branes. The central D5 branes are 
finite in two directions and
represent gauge symmetries in the four dimensional gauge theory. 
Semi-infinite boxes represent global symmetries.
 The intersection of the NS and NS' brane might lead to additional light
states, e.g. D1 strings connecting NS and NS'. 
These states might affect the 4D field theory, but we believe that the
results which are presented in this paper are not influenced by them.
The direction of allowed arrows are North, East and
South-West \cite{HZ}. Outgoing arrows are fundamentals and in-going 
arrows are anti-fundamentals of the gauge group.
Thus Fig.1 represents four dimensional $N=1$ supersymmetric $SU(N)$
with $N_c+X+Y$ fundamentals and $N_c+X+Y$ anti-fundamentals. Notify
that the global symmetry factories into a left and right part. This
is to be
contrasted to other type IIA constructions of N=1 SQCD where only the 
diagonal subgroup of the global flavor symmetry is manifest. Note also
that in this construction $N_f \geq N_c$ (since $X$ and $Y$ cannot be
negative). 

\begin{figure}[H]
  \begin{center}
\mbox{\kern-0.5cm
\epsfig{file=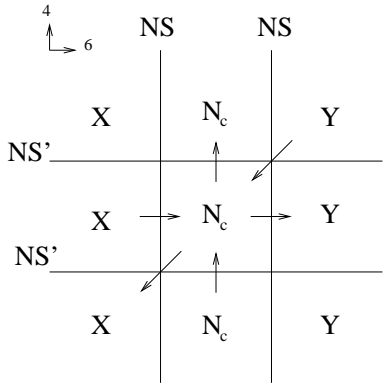,width=5.0true cm,angle=0}}
\label{brane0_fig}
  \end{center}
\caption{$N=1$ Theory with $N_f=N_c+X+Y$ flavors} 
\end{figure}

The construction guarantees that the bending of the four NS
branes is constant along each brane. This is a necessary condition to
ensure that the gauge symmetry is anomaly free \cite{GG}. In addition,
note that the NS' branes have zero bending, since we put the same number
of D5 branes on each side of them. 

The coupling constant of the gauge theory is related to the area of
the middle box. 
\beq
 {1\over g^2} = {\Delta x_4 \Delta x_6 \over g_s l_s^2}
\eeq
The running of the coupling constant is due to the bending of the
branes that surround the middle box. In the type IIA construction of
N=1 gauge theories, one could not define a distance which is directly
related to the running coupling because the D4 branes end on branes
whose world-volume extends in different directions. 
However, since in our construction the two NS' branes do not bend and 
their distance is constant, 
we can relate the beta function to the change of the distance between
the two NS branes. The result is
\beq
b_0 = (X-N_c)  + (Y-N_c)  = N_f-3N_c,
\eeq
which agrees with the field theory result!

\section{N=1 Duality}

In this section we would like to propose a brane motion
that can be related to Seiberg's duality \cite{seiberg}. 
The realization of N=1 Duality using branes in type IIA string 
theory was first demonstrated in \cite{EGK,EGKRS}. Here we show 
that a simple construction in type IIB leads to this duality, too.
\begin{figure}[H]
  \begin{center}
\mbox{\kern-0.5cm
\epsfig{file=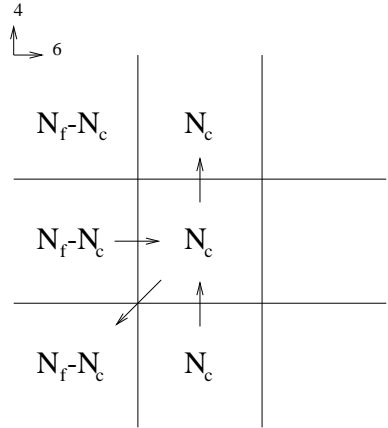,width=5.0true cm,angle=0}}
\label{brane1_fig}
 \end{center}
\caption{The ``electric'' theory. $SU(N_c)$ gauge theory with $N_f$ flavors} 
\end{figure}
Consider the theory in Fig.2 . This theory describes $N=1$ supersymmetric 
$SU(N_c)$ gauge theory with $N_f$ fundamentals and $N_f$ anti-fundamentals.
We refer to this theory as the ``electric" theory. Now attach an additional 
NS brane to the D5 branes at $x_6 = -\infty$. As long as the additional
NS brane is positioned at infinity the theory on the branes remains
unchanged. Now we move the NS brane to a finite 
$x_6$ position. In addition, take the right NS brane and bring it to
$x_6=\infty$. The resulting theory is described in Fig. 3. This theory
is an $SU(N_c)$  theory with $N_f-N_c$ flavors. We refer to this
theory as the "magnetic" theory.
\begin{figure}[H]
  \begin{center}
\mbox{\kern-0.5cm
\epsfig{file=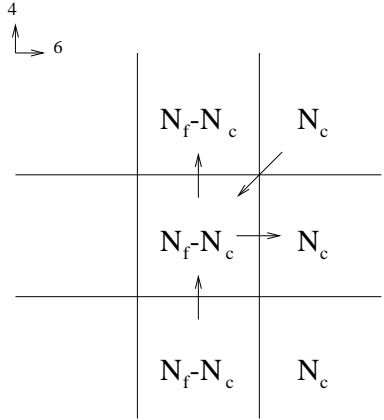,width=5.0true cm,angle=0}}
\label{brane2_fig}
  \end{center}
\caption{The ``magnetic'' theory. $SU(N_f-N_c)$ gauge theory with
  $N_f$ flavors} 
\end{figure}
The intermediate theory (the left and the right NS are at finite
distance in the $x_6$ position) is not dual to either the electric or
the magnetic theories. In fact this is a theory with an $SU(N_c)\times
SU(N_f-N_c)$ gauge group and the following matter content
\begin{table}[H]
\begin{displaymath}
\begin{array}{l c@{ }c}

 & \multicolumn{1}{c@{\times}}{SU(N_c)}
 & \multicolumn{1}{c@ {}}{SU(N_f-N_c)} \\ 
\hline
 &  N_f\Yfund & 1 \\  
 &  N_c \overline{\Yfund} & 1  \\
 & 1 & (N_f-N_c)  \Yfund\\ 
 & 1 & N_f \overline{\Yfund} \\
 & \overline{\Yfund} & \Yfund
\end{array}
\end{displaymath}
\caption{The matter content of the 'intermediate' theory}
\label{table:intermideate}
\end{table}

However, the magnetic and the electric theories can be viewed as two limits
of this intermediate theory.

While we cannot justify the motion of the branes which leads to the duality,
it reproduces the known field theory results. 

As a simple test of the above suggestion, we will prove that the
relation between giving a mass terms in the electric
theory and Higgsing in the magnetic theory (and vice versa)
holds in the
brane configuration. Adding a mass term for one of the quarks 
in the electric theory 
reduces the flavor symmetry by one. In the magnetic theory this 
corresponds to giving a mass to one of the mesons which via the
equations of motion induces a vev for one of the magnetic quarks   
reducing both colors and flavors by one unit.

This correspondence can be seen in the brane construction in the 
following way: reconnect the D5 brane in the left-upper box 
of Fig.2 to the two other D5 branes in the left column. Now the D5
brane is infinite in the $x_4$ direction and can be lifted in the 
$x_5$ direction. The $x_5$ position is proportional to the mass term.
The result is an electric theory with $N_f-1$ flavors. Introducing the
additional NS brane and moving the right NS brane, we obtain the
magnetic theory in which the middle column of Fig.3 has $N_f-N_c-1$
D5 branes. The $x_5$ position is related to the mass term for 
the meson. This magnetic theory has $N_f-N_c-1$ colors and $N_f-1$ 
flavors, as expected. 

In the same way we can show the relation between Higgsing in 
the electric theory and mass terms in the magnetic theory. 
In this case we reconnect D5 branes in the middle column of fig.2 to 
obtain one D5 brane that is infinite in the $x_4$ direction but 
finite in the $x_6$ direction and lift it in the $x_5$ direction.
This reduces color and flavor by one and corresponds to Higgsing 
in the electric theory where the $x_5$ position is proportional to
$\langle Q \tilde Q \rangle$. After the brane motion we obtain the
magnetic theory with flavor reduced by one which is related 
to giving a mass term to one of the quarks in the magnetic
theory.

In the above discussion, we did not consider the superpotential in the
two dual descriptions. 
Duality of the field theories requires zero superpotential in the
electric theory  and a superpotential of the form $W=Mq\tilde q$ in the 
magnetic theory. Since in
our description the flavor symmetry boxes are semi-infinite, the
mesons in both sides are not dynamical, in contrast to the
situation in the magnetic theory in field-theory. This 
obstacle can probably be solved by adding an extra NS brane at large
value of $x_6$ in the magnetic brane configuration. We leave this
issue as an open question. 

 From the field theory point of view it looks like two different
  degenerations of a field theory with a product gauge group. By
  decoupling one of the factors we turned a local symmetry into global
  flavor symmetry. One also would like to
  check that the moduli spaces in the electric and magnetic theory
  match. We have given some 
  evidence for this by relating massterms and Higgsing in the dual
  theories. Maybe the introduction of D7 branes is necessary (similar
  to the use of D6 branes in type IIA) to introduce matter to make
  this match complete. In the
  type IIA brane construction of Seiberg's duality\cite{EGK} a certain brane
  creation process was a crucial ingredient. In this approach 
  the moduli space
  could be matched precisely and the meson was identified. But this 
  approach
  also has disadvantages e.g. only a diagonal subgroup of the flavor
  symmetry is manifest and an  Fayet-Ilioupoulos term had to be
  introduced to avoid singularities in the brane motion. Such a
  singularity was never a problem the type IIB approach. In summary
  it must be said that any brane construction of Seiberg's duality 
  presented so far captures many features of the field theory duality but 
  has also some problems. 
\section{Anomaly cancelation in chiral theories}

In this section we study $SU(N_c)$ gauge theories with an
symmetric/anti-symmetric tensor, $N_f$ fundamentals and $\bar N_f$
anti-fundamentals. Anomaly cancelation requires $\bar N_f - N_f =
N_c+4$ (or $\bar N_f - N_f = N_c-4$) for theories including the
symmetric (or anti-symmetric) representation.

We cannot derive this result from string theory but the 
configuration is fixed by the requirement that NS branes 
have constant bending and that the field theory anomaly is 
canceled. The relevant brane configuration for the 
symmetric case is described in
Fig.4 
\begin{figure}[H]
  \begin{center}
\mbox{\kern-0.5cm
\epsfig{file=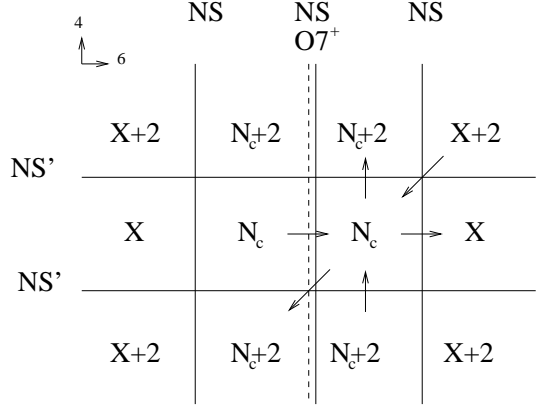,width=7.0true cm,angle=0}}
\label{brane3_fig}
  \end{center}
\caption{$N=1$ $SU(N)$ theory with symmetric, fundamentals
  and anti-fundamentals. The dashed line denote the $O7^+$.} 
\end{figure} 
In addition to the three  NS branes, the two NS' branes and the D5
branes we added on top of the middle NS brane an 
orientifold-sevenplane 
with world volume 01234789. The presence of the
orientifold induces a $Z_2$ symmetry with respect
 to the $x_5$ and $x_6$
directions. An orientifold plane with RR charge $-8$ ($+8$) will be
denoted by $O7^-$ ($O7^+$) and gives rise to a $SU(N)$ gauge theory
with an anti-symmetric (symmetric) representation.

In order to ensure that the bending of each NS brane remains  
constant along the brane, we added additional D5 branes 
around the $N_c$ D5 branes in the middle. The 
additional branes are responsible for extra matter in the
fundamental and anti-fundamental of the gauge group.
Above the upper NS' brane there should be 2 extra D5 brane at 
each side of the mirror (and similarly two extra D5 branes 
below the lower NS', at each side), to cancel the chiral anomaly
in field theory. Although we have no string theory argument for this
rule, it should be possible to derive it directly from string theory.
Note that it is not sufficient to require a constant bending. There
should be two extra D5 branes above and below the middle row. Any
other number of D5 branes leads to an anomaly in the field theory.
We assume that this effect is due to the presence of the orientifold plane.   

Let us count the number of fundamentals/anti-fundamentals which
interact with the color group. The in-going arrows are
anti-fundamentals and the outgoing arrows are fundamentals. There is
one exception: The South-West arrow which crosses the $O7^+$ plane is
counted as anti-fundamental and not as fundamental. The argument
for this is the following: A superpotential term $A \tilde Q
\tilde Q$ is formed by the symmetric
(the horizontal arrow which connect the middle box to its mirror), the
anti-fundamentals (the vertical arrow which connects the lower box to
the middle box in the mirror) and the South-West arrow. Since the
superpotential is a scalar, the diagonal arrow should represent
anti-fundamentals. Hence we have $\bar N_f = 2N_c + X+6$ and
$N_f = N_c + X + 2$. 

We conclude that 
\beq
\bar N_f - N_f = N_c + 4 ~.
\eeq
Similarly, in the case of anti-symmetric matter and $O7$ plane we
obtain
\beq
\bar N_f - N_f = N_c - 4 ~.
\eeq        

We can also calculate, approximately, the beta function of these
theories. As we explained in section 2, the relation between the
one-loop beta function and the bending of the branes, is meaningful
only when two of the branes that surround the gauge group box remain
parallel. In the current configuration this is not the case, since the
two NS' branes are not parallel. The D5
and the $O7^+$ bend the NS' brane in different directions and
therefore the NS'  can never be made 
flat in all directions.
However, if the bending of the right NS brane is much larger than the
bending of the NS' brane, we expect to get an approximate answer.
By approximate we mean that the dependence on $N_c$ and $N_f$ is correct.

The bending of the right NS is $X-N_c$, hence
\beq
b_0\approx N_f-2N_c-2
\eeq
Note that the field theory result is
\beq
b_0= N_f-2N_c+3
\eeq
Which indeed agrees when $2N_c-N_f \gg 1$. Similarly, the approximate
value of the beta function can be calculated in the case of the chiral
theory with anti-symmetric matter. 

\section{SO and Sp Gauge Groups}

We would like to comment on the one-loop beta function in the case of
$N=1$ SQCD with gauge groups $SO(N_c)$ and $Sp(N_c)$ (with 
$N_c$ even). The realization of these models was worked out in \cite{HZ} 
and it is similar to the theory in Fig.4. The brane configuration for
$SO(N_c)$ is given in Fig.5
\begin{figure}[H]
  \begin{center}
\mbox{\kern-0.5cm
\epsfig{file=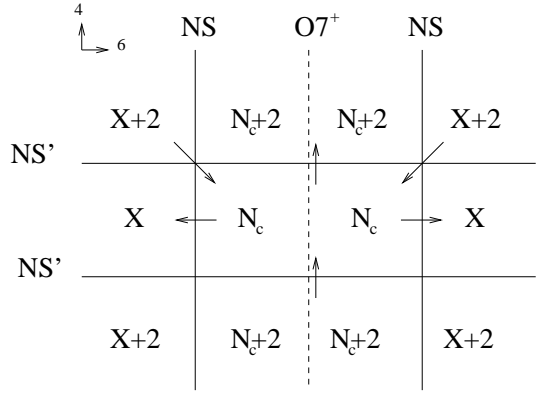,width=7.0true cm,angle=0}}
\label{brane4_fig}
  \end{center}
\caption{$N=1$ $SO(N_c)$ theory. The dashed line denote the $O7^+$.} 
\end{figure} 
Note that in contrast to the previous case, there is no NS brane at
$x_6=0$ and the color box should be understood as one box. 
Here we assume that the orientifold plane imposes the same 
constraints on the number of branes in the upper and lower row as
in the previous section. This brane rule can be supported by field
theory arguments. Chiral theories and orthogonal/symplectic theory are
closely related by duality\cite{Berkooz}. A theory with
$Sp(N_c-4)\times SU(N_c)$ gauge group can be constructed in a similar
way to the theory in Fig.5, but with an $O7$ plane and additional NS
branes at finite $x_6$ and $-x_6$. By moving the middle NS brane
and its mirror towards the orientifold plane and moving one of them to
$x_7=\infty$ we obtain an $SU(N_c)$ theory with antisymmetric tensor
and extra matter.
This process was already described in ref.\cite{HZ}. Here we conclude that
two extra D5 branes above and below the gauge box in
the chiral case (Fig.4), are also necessary in the orthogonal and
symplectic cases. 

According to the rules of the previous section, we will not be able to
get the accurate beta function of these theories since a completely
flat NS' brane cannot be achieved.
But it is again possible to get the correct dependence on $N_f$ and $N_c$. 
The number of flavors in this case is 
\beq
2N_f= 4X+2N_c+8
\eeq
The one-loop beta function is obtained, approximately, by $2X-2N_c$
\beq
b_0 \approx N_f-3N_c-4,
\eeq
in agreement with the field theory result
\beq
 b_0 = N_f-3N_c+6.
\eeq
A similar result can be obtained for the $Sp$ case. In this case we
replace the $O7^+$ by $O7$, and we again capture the $N_f-3N_c$ 
dependence.

\section{Acknowledgments}

We thank Nissan Itzhaki, Udi Fuchs and especially Shimon
Yankielowicz for discussions.

\end{document}